\documentclass[superscriptaddress,twocolumn,showpacs,prb,floatfix]{revtex4}
\usepackage[utf8]{inputenc} 
\usepackage{amsmath}
\usepackage{braket}
\usepackage{graphicx}
\usepackage{amsfonts}
\usepackage{pgfplots}
\usepackage{csquotes}
\usepackage{hhline}
\usepackage{amssymb}
\usepackage{listings}
\usepackage{color}
\makeatletter
\newcommand{\ssymbol}[1]{^{\@fnsymbol{#1}}}
\makeatother
\definecolor{codegreen}{rgb}{0,0.6,0}
\definecolor{codegray}{rgb}{0.5,0.5,0.5}
\definecolor{codepurple}{rgb}{0.58,0,0.82}
\definecolor{backcolour}{rgb}{0.95,0.95,0.92}
 
\lstdefinestyle{mystyle}{
    backgroundcolor=\color{backcolour},   
    commentstyle=\color{codegreen},
    keywordstyle=\color{magenta},
    numberstyle=\tiny\color{codegray},
    stringstyle=\color{codepurple},
    basicstyle=\footnotesize,
    breakatwhitespace=false,         
    breaklines=true,                 
    captionpos=b,                    
    keepspaces=true,                 
    numbers=left,                    
    numbersep=5pt,                  
    showspaces=false,                
    showstringspaces=false,
    showtabs=false,                  
    tabsize=2
}
 
\usepackage{subfigure}

\usepackage{dcolumn}
\usepackage{tabularx}
\usepackage[colorlinks=true,linkcolor=blue,citecolor=blue,urlcolor=blue]{hyperref}
\usepackage{longtable}
\usepackage{braket}
\usepackage{float}
\newcolumntype{C}{>{\centering\arraybackslash}X}
\pgfplotsset{compat=1.17}
\begin{document}
\title{Experimental realization of BB84 protocol with different phase gates and SARG04 protocol}

\author{Sinchan Ghosh}
\email{sinchanghosh008@gmail.com}
\affiliation{Department of Physics, Ramakrishna Mission Vidyamandira (Affiliated to Calcutta University), Belur, Howrah, 711202, West Bengal, India}
\author{Harsh Mishra}
\email{harshm19@iiserbpr.ac.in}
\affiliation{Department of Physical Sciences, Indian Institute of Science Education and Research, Berhampur, India}

\author{Bikash K. Behera}
\email{bikash@bikashsquantum.com}
\affiliation{Bikash's Quantum (OPC) Pvt. Ltd., Balindi, Mohanpur, 741246, Nadia, West Bengal, India}
\author{Prasanta K. Panigrahi}
\email{pprasanta@iiserkol.ac.in}
\affiliation{Department of Physical Sciences,\\ Indian Institute of Science Education and Research Kolkata, Mohanpur 741246, West Bengal, India}

\begin{abstract}
Cryptography in the modern era is very important to prevent a cyber attack, as the world tends to be more and more digitalized. Classical cryptographic protocols mainly depend on the mathematical complicacy of encoding functions and the shared key, like RSA protocol in which security depends upon the fact that factoring a big number is a hard problem to the current computers. This means that high computing power can help you crack traditional encryption methods. Quantum machines claim to have this kind of power in many instances. Factorization of big numbers may be possible with Shor's algorithm with quantum machines in considerable time. Apart from this, the main problem is key sharing i.e., how to securely share the key the first time to validate the encryption. Here comes quantum key distribution. Two parties who are interested in communication with each other, create a process, which claims considerable security against an eavesdropper, by encoding and decoding information in quantum states to construct and share a secret key. Quantum key distribution may be done in a variety of ways. This paper begins with experimental verification of the BB84 procedure utilizing four bases (using phase gates) followed by the experimental realization of the SARG04 protocol which was derived from BB84 Protocol to overcome PNS attack. The possibility of a third-party attack and the effect of noise is considered and implemented. The IBM Quantum Experience platform was used for all of the implementations.
\end{abstract}

\maketitle
\section{Introduction}
\label{SecI}
Quantum cryptography claims considerable attention by promising top-notch security by allowing back-end operations that exploit the weirdness of quantum physics\cite{RQC2011Science2011}. Over time, this field has been developed positively with specialization in the security of communication. The idea of cryptography using quantum physics was initiated about 27 years ago. The first quantum key distribution protocol was invented to develop a secure construction of a secret key despite the presence of a third-party hacking adversary i.e. Eve, withholding maximum access to technology. This is naturally done to make sure that the developed protocol's security is not compromised \cite{qkd_PathakCRC2018}. This became a new emergence in quantum information theory, and progress has continued to accelerate since then. This first protocol was known as the BB84 protocol \cite{BBQCIEEE1984, BCBTCS2014} introduced by Bennett and Brassard in the year 1984. Furthermore, several protocols have been developed, over the years, to make sure the key distribution allows security from eavesdropping at the maximum possible level. Quantum Key Distribution started developing over time, both in terms of efficiency and experimentation \cite{HFPR2014, BFO2017, AFPRA2012, CHBJoC1992, MMNP2019, DAMIEEE2013}. SARG04 protocol was built to make BB84 more secure by decreasing PNS attacks. Experimenting with these protocols on IBM Quantum Experience (IBM QX) using local simulators and actual quantum processors is crucial because it helps to understand the importance of gate operations on qubits in quantum computation. Also, the simulation of a circuit from the real quantum device (up to a certain limit) gives an idea of real-time performance and errors and helps to apply appropriate implementation techniques. A part of reference \cite{DAMIEEE2013} implements BB84 protocol using 8 qubits on the IBM QX platform with and without the presence of Eavesdropper, using two bases and also a part of the reference \cite{ERQKD2020} implements BB84 protocol on the IBM QX platform using 8 qubits with and without the presence of Eavesdropper using three bases. To emphasize the relevance of phase gates in quantum computation, we have investigated implementing the BB84 protocol utilizing four bases (using Phase gates) as well as the SARG04 protocol on the IBM QX platform.

\section{BB84 PROTOCOL WITH FOUR BASES}
\label{SecII}
BB84 protocol was designed such that every bit of randomly chosen string would be encoded into the two polarization states of a single photon (which can be represented by qubits). As photon is a quantum particle, without destroying the photon, its polarization state cannot be measured, which implies that the encoded information in the photon is frangible \cite{QKDL12}. Considering this fact, we assume that this is the reason why an eavesdropping attack does not allow the information to be available to the eavesdropper. The BB84 protocol can be briefly understood by considering the following points \cite{NCCUP2000}. Implementation of BB84 protocol on IBM QX using four bases has been performed in the first part of this paper.

\begin{itemize}
    \item \raggedright Alice produces a bit-string, randomly choosing a set of 2n (say) data bits
    \item  She then encodes her data bit using four states (two basis X and Z) $\Ket{0}$,$\Ket{1}$,$\Ket{+}$ and $\ket{-}$ and sends it to Bob. 
    \item Bob receives the qubits and announces that fact. He then measures the received qubits on the X or Z basis at random.
    \item Then both announce publicly the basis on which they have measured.  After that, they find n (say) bits are encoded and decoded using the same basis. Those are left for the preparation of the secret key.
    \item An eavesdropper, Eve can attack while transmitting. She measures the qubits in the X or Z basis at random, then encodes that qubit again on the same basis. If Eve's decoding base matches with Alice's encoding base, then Eve gets the right information. 
    \item Now they need to check whether or not eavesdropping took place. So, n/2 bits serve as a check on the intervention and are shared with Bob through a classical channel by Alice. 
    \item Now Bob compares his measurement of those n/2 qubits with Alice's preparation. If the comparison result does not show an acceptable number of consistent bits despite having chosen the same basis, then the protocol fails. The existence of an eavesdropper or noise in the quantum channel is certain. Now they leave that channel and try for another channel.
    \item Now, if that comparison result shows an acceptable number of consistent bits, then they can continue and proceed for communication using the remaining n/2 bits as the secret key.
\end{itemize}
We aim to increase the efficiency of the protocol by decreasing the amount of bit information revealed to the eavesdropper. If any accepted key bit is measured by Eve with the same basis then the attack will remain untraced. So, using four basis states instead of two decreases the probability of Eve measuring the qubit with the same basis as used by Alice for encoding. So, there is an enhance in security.
\subsection{IMPLEMENTATION OF BB84 PROTOCOLS ON IBM QX}
\label{SecIIA}

This section portrays the experimental implementation of the BB84 protocol using four bases. Let's say the two parties, Alice and Bob want to share a secret key through a quantum protocol. Thus, let's say Alice chooses 2n number of random data bits and sends it to Bob. For the encoding procedure, she randomly uses any one of the four basis and sends the resultant state to Bob through a public quantum channel. X \& Y basis are the two of the four bases. Among the other two, one basis is formed applying hadamard \& T gate consecutively and the other is formed applying hadamard \& Z gate consecutively. We have discussed the operations of these gates later. After Bob receives the 2n qubits, he measures each of those qubits on one of the four basis at random. Alice and Bob henceforth crosscheck their basis measurements. The classical post-processing part which has to be followed is the same as done for the implementation of BB84 protocol using two bases. Let's take a look at the procedure of implementation. 

\begin{figure}[H]
    \centering
    \includegraphics[scale=0.3]{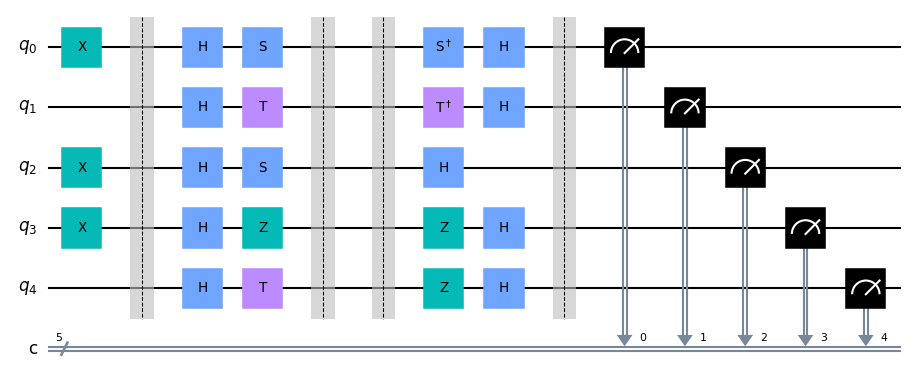}
    \caption{Implementation of BB84 protocol using four bases. The left part of the first barrier portrays the initial bit-string. The part between the first \& second barrier and the part between the second \& third barrier portray Alice's encoding and Bob's decoding respectively. The distance between the second and third barrier can be considered as the quantum channel. The right part of the fourth barrier contains Bob's measurements.}
    \label{SPcircuit}
\end{figure}

\begin{table}[ht]
    \centering
    \begin{tabular}{|c|c|c|c|c|c|}
    \hline\hline\\[0.5ex]
    Qubit Index & q[0] & q[1] & q[2] & q[3] & q[4] \\ [0.5ex]
    \hline
    Alice & 1 & 0 & 1 & 1 & 0 \\ [0.5ex]
    \hline
    Alice's Basis & Y & HT & Y & HZ & HT\\ [0.5ex]
    \hline
    Bob's Basis & Y & HT & X & HZ & HZ \\ [0.5ex]
    \hline
    Result & A & A & D & A & D \\ [0.5ex]
    \hline 
    \end{tabular}
    \caption{Initial value of all the qubits i.e. from q[0] to q[4] is $\Ket{0}$ as default. We have used not gate to get qubit value $\Ket{1}$. The qubits which are measured with the same basis as used by Alice, are the ones that are considered to be key bit. A stands for ``Accepted" whereas D stands for ``Discarded".}
    \label{qkdtable1}
\end{table}

Table.\ref{qkdtable1} shows the qubits which will be taken into consideration whereas Table.\ref{qkdt2} shows gates that are required to measure along with different bases. We use here phase gates of different angles. A phase gate applies a phase of $e^{i\theta}$ to the $\Ket{1}$ state of the qubit. H is the hadamard gate while S is the phase gate for angle $\pi/2$. T is the phase gate for angle $\pi/4$. And the Z gate is the phase gate for angle $\pi$. The explicit matrix forms of phase gates and hadamard gate are given below. S$\ssymbol{2}$, T$\ssymbol{2}$, Z are the complex conjugate forms of S, T, and Z gates respectively. Expected results based on the theory are shown in Table.\ref{qkdtable3}. 

\[ S= \left( \begin{array}{cc}
 1 & 0\\
 0 & e^{-i\pi/2}
\end{array} \right)
 \quad H = \frac{1}{\sqrt{2}}
\left( \begin{array}{cc}
  1 & 1\\
  1 & -1
\end{array} \right)
\]

\[ T= \left( \begin{array}{cc}
    1 & 0\\
    0 & e^{i\pi/4}
\end{array} \right)
 \quad Z =
\left( \begin{array}{cc}
  1 & 0\\
  0 & -1
\end{array} \right)
\]

\begin{table}[ht]
    \centering
    \begin{tabular}{|c|c|c|c|c|c|}
    \hline\hline\\[0.5ex]
    Qubit Index & q[0] & q[1] & q[2] & q[3] & q[4] \\ [0.5ex]
    \hline
    Alice's gates for & X & - & X & X & - \\ string construction & & & & &  \\ [0.5ex]
    \hline
    Alice's gates for & HS & HT & HS & HZ & HT \\ string encoding & & & & & \\ [0.5ex]
    \hline
    Bob's decoding & $S\ssymbol{2}$H & $T\ssymbol{2}$H & H & ZH & ZH \\ gates & & & & & \\ [0.5ex]
    \hline 
    \end{tabular}
    \caption{Gate utilization for the example of 5-qubit model. Since S is a hermitian matrix, $SS\ssymbol{2}=I$ which implies Bob needs to use $S\ssymbol{2}H$ gates to invert the operation of HS gates (i.e., decoding information). This is similar for T gate.}
    \label{qkdt2}
\end{table}

\begin{table}[ht]
    \centering
    \begin{tabular}{|c|c|}
    \hline\hline
    Qubit Index & Expected\\ & Result  \\ [0.5ex]
    \hline
    q[0] & 1 (100\%)\\[0.5ex]
    \hline
    q[1] & 0 (100\%) \\[0.5ex]
    \hline
    q[2] & 0 (50\%) \& 1 (50\%) \\[0.5ex]
    \hline
    q[3] & 1 (50\%) \\[0.5ex]
    \hline
    q[4] & 0 (50\%) \& 1 (50\%) \\[0.5ex]
     \hline
    \end{tabular}
    \caption{These are the expected results based on the BB84 protocol using four bases. Parentheses include the probability of getting that particular bit after measurement.}
    \label{qkdtable3}
\end{table}

\subsubsection{Simulation using local simulator}
\label{SecIIAi} 

\begin{figure}[H]
    \centering
    \includegraphics[scale=0.25]{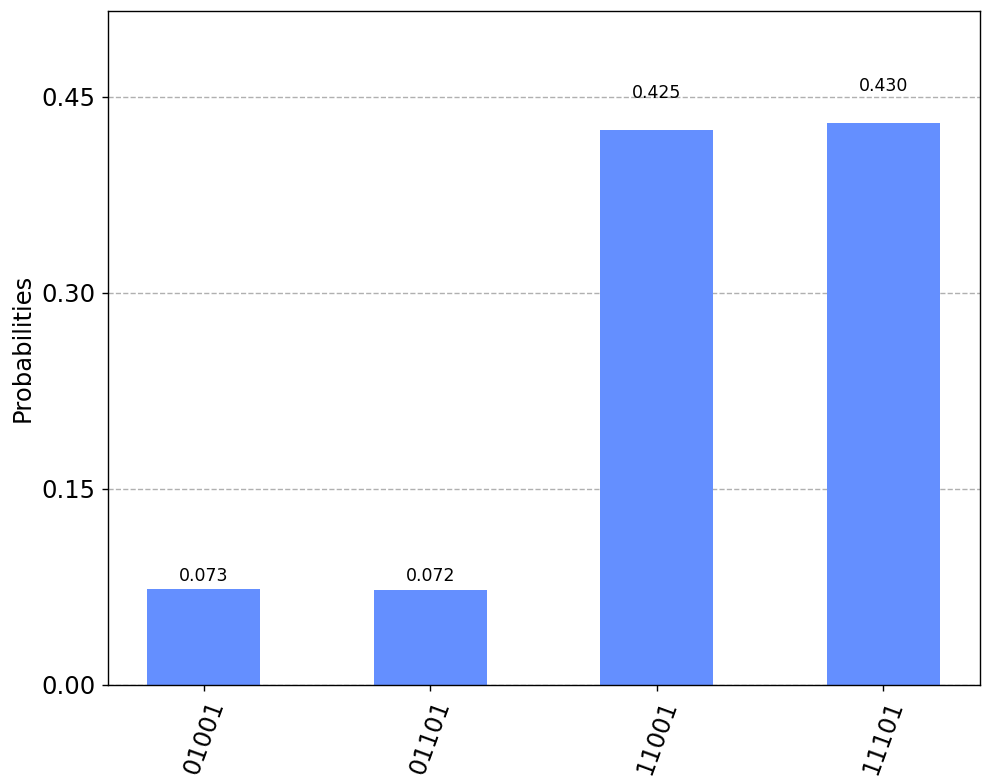}
    \caption{Result of implementation of BB84 protocol using four bases acquired through IBM QX (ibmq\_qasm\_simulator) using 8192 runs. In all of the outputs, the value of q[0] is 1 which implies that the probability of acquiring 1 is 100\% which is what we expected. Similarly, the probability of measuring other qubits can be calculated like this and will be similar to the result in table \ref{qkdtable3}.}
    \label{SPhistogram}
\end{figure}

The implementation for this 5-qubit example is given in Fig.\ref{SPcircuit} and the results of this implementation through local simulator are given in Fig.\ref{SPhistogram}. This local simulator is a classical machine that approximates quantum calculation up to a limit. So, this implementation result is similar to the theoretically expected result in Table.\ref{qkdtable3}. The probability of q[0] through experimental results is calculated as shown in the reference paper \cite{MKAAANICCAIS2019}. Table.\ref{qkdtable5} shows the experimental results after implementation.

\subsubsection{Simulation using real quantum device(ibmq\_manila)}
\label{SecIIAii}

\begin{figure}[H]
    \centering
    \includegraphics[scale=0.23]{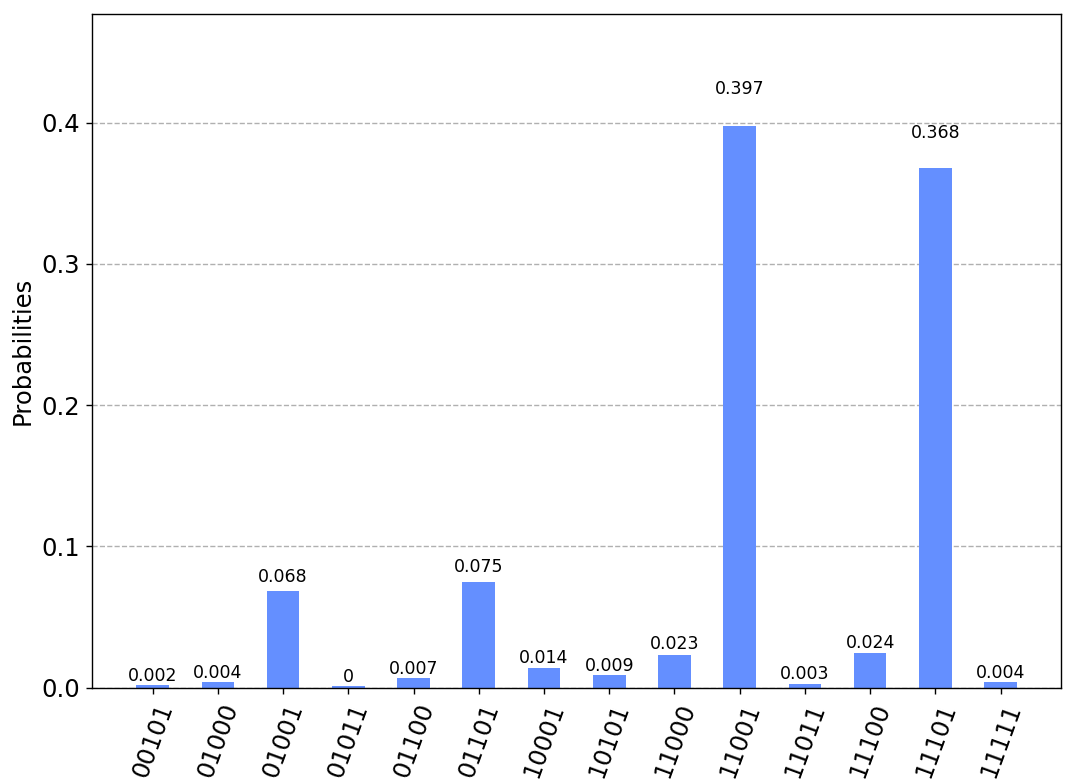}
    \caption{Result of implementation of BB84 protocol using four bases acquired through IBM QX (ibmq\_manila) using 8192 runs. In all of the outputs, the value of q[0] is not 1 which implies that the probability of acquiring 1 is less than expected 100\%. The deviation is due to various computational error in real device. Similarly, the probability of measuring other qubits are calculated like this and tabulated in Table.\ref{qkdtable5}.}
    \label{SPhistReal}
\end{figure}

This section portrays the experimental implementation of the BB84 protocol using four bases in a real quantum device through IBM QX. The back-end device used here is ibmq\_manila, a five qubit processor. As simulated from real quantum device, the output result (in the Fig.\ref{SPhistReal}) is a bit deviated from the expected result (in the Fig.\ref{SPhistogram}) due to the various computational error. Result for every qubit acquired from the real device is shown in Table.\ref{qkdtable5}.

\begin{table}[ht]
    \centering
    \begin{tabular}{|c|c|c|}
    \hline\hline
    Qubit Index & Experimental\\ & Result \\ [0.5ex]
    \hline
    q[0] & 0 (3.9\%) \& 1 (96.1\%)\\[0.5ex]
    \hline
    q[1] & 0 (99.1\%) \& 1 (0.9\%)\\[0.5ex]
    \hline
    q[2] & 0 (48.8\%) \& 1 (51.2\%) \\[0.5ex]
    \hline
    q[3] & 0 (3.0\%) \& 1 (97.0\%) \\[0.5ex]
    \hline
    q[4] & 0 (14.4\%) \& 1 (85.6\%) \\[0.5ex]
    \hline
    \end{tabular}
    \caption{These are the experimental results for every qubit obtained by simulating BB84 protocol using four bases in IBM QX (ibmq\_manila) using 8192 runs. Parentheses include the probability of getting that particular bit after measurement.}
    \label{qkdtable5}
\end{table}

\subsubsection{Fidelity calculation of individual qubits}
\label{SecIIAiii}
The fidelity of a quantum state is a measure of closeness. Here, we measure fidelity between the state of qubit before encoding by Alice and the state after decoding by Bob. If the fidelity value is close to 1, we can take the bit as a key bit. We have calculated the fidelity separately for 5 qubits using the equation below\cite{Jozsa1994}-
\begin{equation}
\textrm{Fidelity }= tr(\sqrt{\rho^{(\text{E})}\rho^{(\text{T})}\rho^{(\text{E})}})
\end{equation}
where  theoretical density matrix $\rho^{(\text{T})}$ is simply the difference of the probabilities of 0 and 1 while experimentally obtained density matrix $\rho^{(\text{E})}$ is calculated as\cite{BKbehera}-
\begin{equation}
\rho^{(\text{E})}=\frac{1}{2}(I+\langle X\rangle X +\langle Y\rangle Y+\langle Z\rangle Z)
\end{equation}

The fidelity value of each separate qubit of BB84 protocol is tabulated in  Table.\ref{fid_table} below. Qubits with fidelity value very close to 1, are taken as key bit to construct the key of encryption.
\begin{table}[ht]
    \centering
    \begin{tabular}{|c|c|}
    \hline
    \hline
    Qubit Index & Fidelity \\[0.5ex]
    \hline
    q[0] & 0.9903666531616115  \\[0.5ex]
    \hline
    q[1] & 0.9998826058435594  \\[0.5ex]
    \hline
    q[2] & 0.5105520995810717  \\[0.5ex]
    \hline
    q[3] & 0.9988745629501058 \\[0.5ex]
    \hline
    q[4] & 0.21158200569024022  \\[0.5ex]
    \hline
    \end{tabular}
    \caption{Fidelity values for individual qubits of BB84 protocol circuit calculated based on the result acquired from ibmq\_manila on IBM QX.}
    \label{fid_table}
\end{table}

\subsubsection{Error mitigation in real quantum device}
\label{SecIIAiv}

\begin{figure}[H]
    \centering
    \includegraphics[scale=0.12]{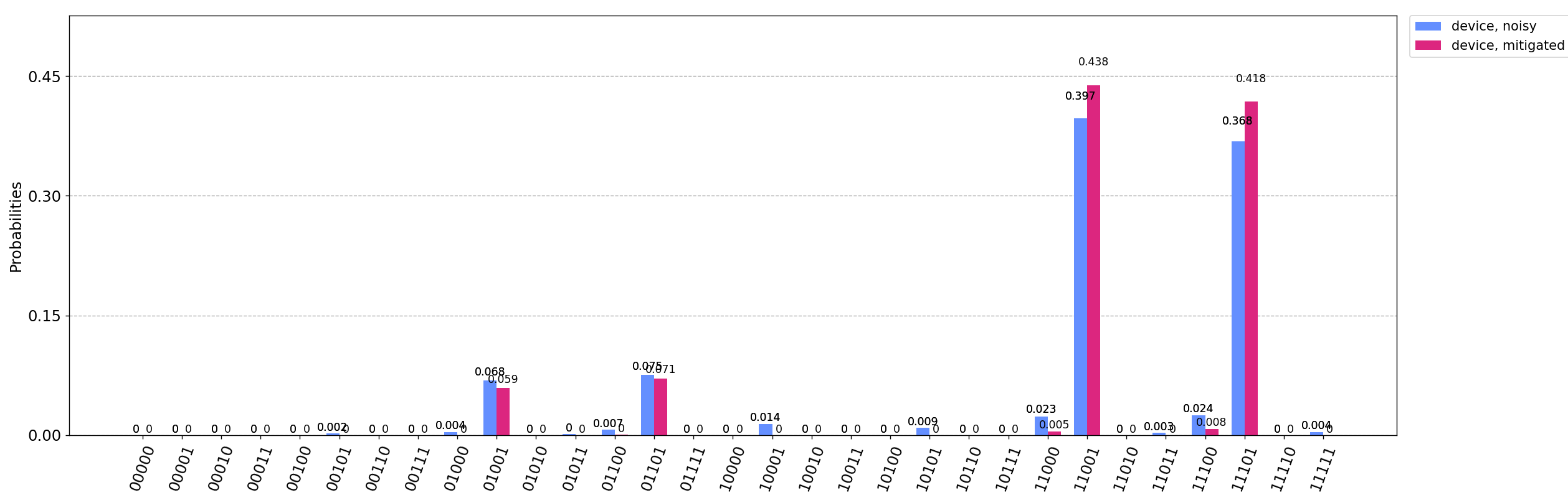}
    \caption{Result of implementation of BB84 protocol using four bases acquired through IBM QX (ibmq\_manila) using 8192 runs after processing an error mitigation. The result before and after mitigation is plotted simultaenously. Clearly, mitigated result is more close to accurate result \ref{qkdtable3}.}
    \label{mitigationhist}
 
\end{figure}

 The effect of different types of noise, that occur throughout the computation and make the output quite noisy, is very complex and needs to consider each gate transformation. However, the noise occurring in the final measurement can be reduced by the error mitigation technique. The idea behind the error mitigation process is that the exact noiseless outcomes are assumed from the noisy result of the real device. Then those results are used to mitigate the errors for a more general form of state. Our BB84 circuit is processed by such an error mitigation process\cite{GEM2020}. The histogram in Fig.\ref{mitigationhist} indicates the reduction of noise while computation.
\subsection{DIFFERENT ATTACKS ON THE QUANTUM CIRCUIT}
\label{SecIIB}
\subsubsection{Third party interpretation}
\label{SecIIBi}

\begin{figure}[H]
    \centering
    \includegraphics[scale=0.25]{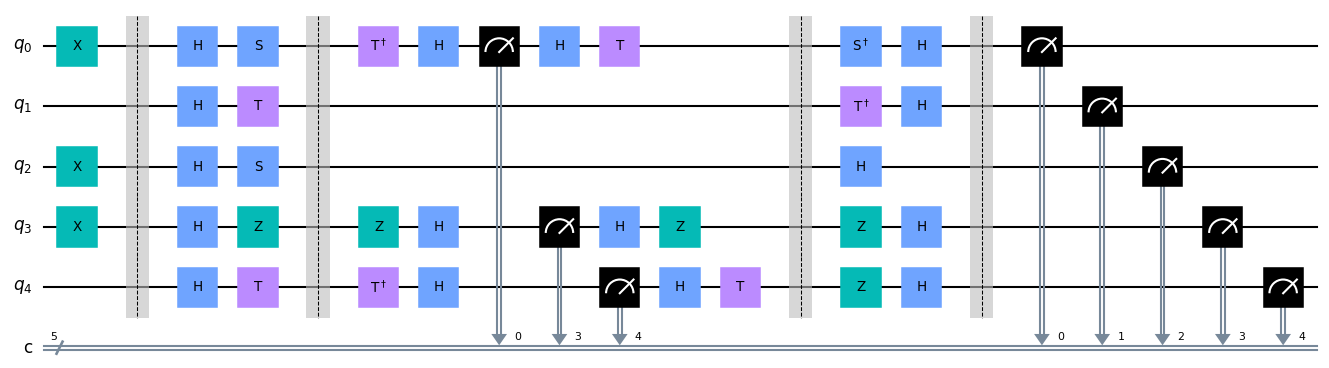}
    \caption{Eve decodes in some assumed basis, measures, and then again encodes the qubits while Alice's encoded information is being sent to Bob through the channel. The qubits cannot be copied as per the No-Cloning theorem \cite{qkd_WoottersNature1982}. Let's say Eve attacks on q[0], q[3], q[4].}
    \label{SPAttack}
\end{figure}

\begin{figure}[H]
    \centering
    \includegraphics[scale=0.3]{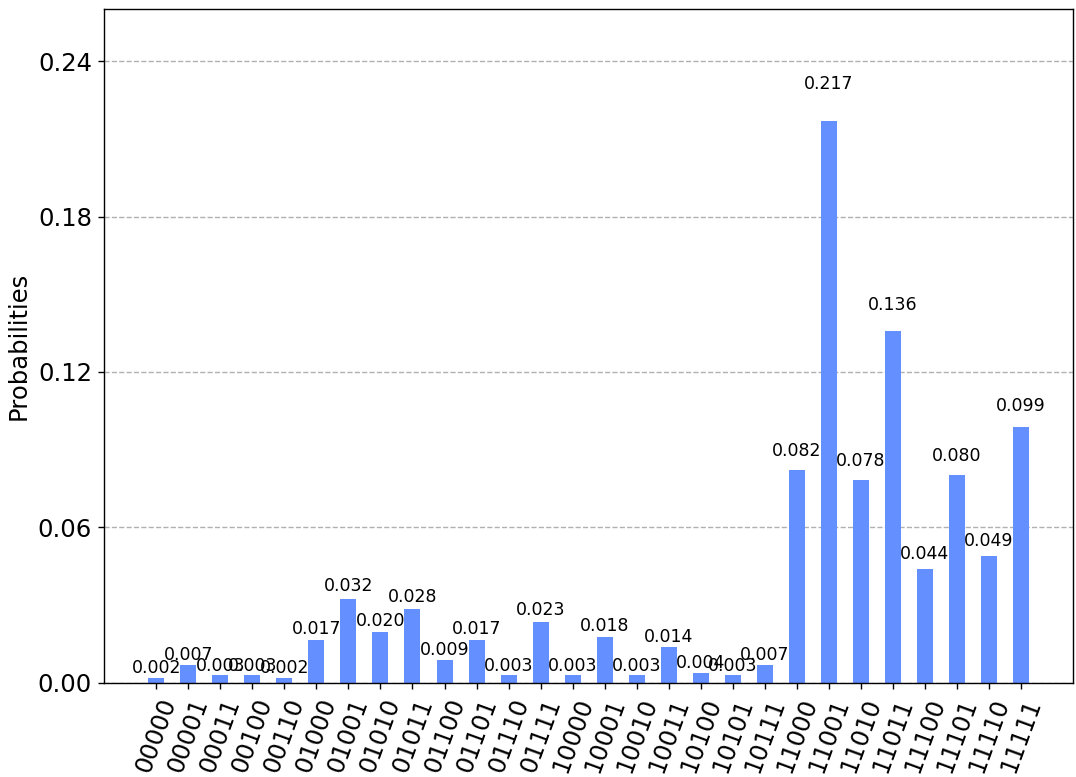}
    \caption{Results for implementation of an eavesdropper's attack on BB84 protocol using 5 qubits on IBM QX (ibmq\_manila). Eve measures q[1], q[3], q[4] out of which she gets q[3] right. Once Bob reveals his measurements to Alice, they would realize that eavesdropping has happened as the outputs do not tally with Alice's preparation. Calculating the probability outcomes for the three attacked qubits, we get the results as given in the table \ref{qkdtable6}.}
    \label{SPAttackhist}
\end{figure}

We now move forward with implementing the circuit which portrays a third party interpretation i.e., an eavesdropper trying to attack the circuit in the quantum channel. The circuit is shown in Fig.\ref{SPAttack} whereas the results of the implementation are in Fig.\ref{SPAttackhist}. Table.\ref{qkdtable5} shows the probability of occurrence of each qubit for Bob when eavesdropping takes place.
After Error Mitigation: 
\begin{figure}[H]
    \centering
    \includegraphics[scale=0.12]{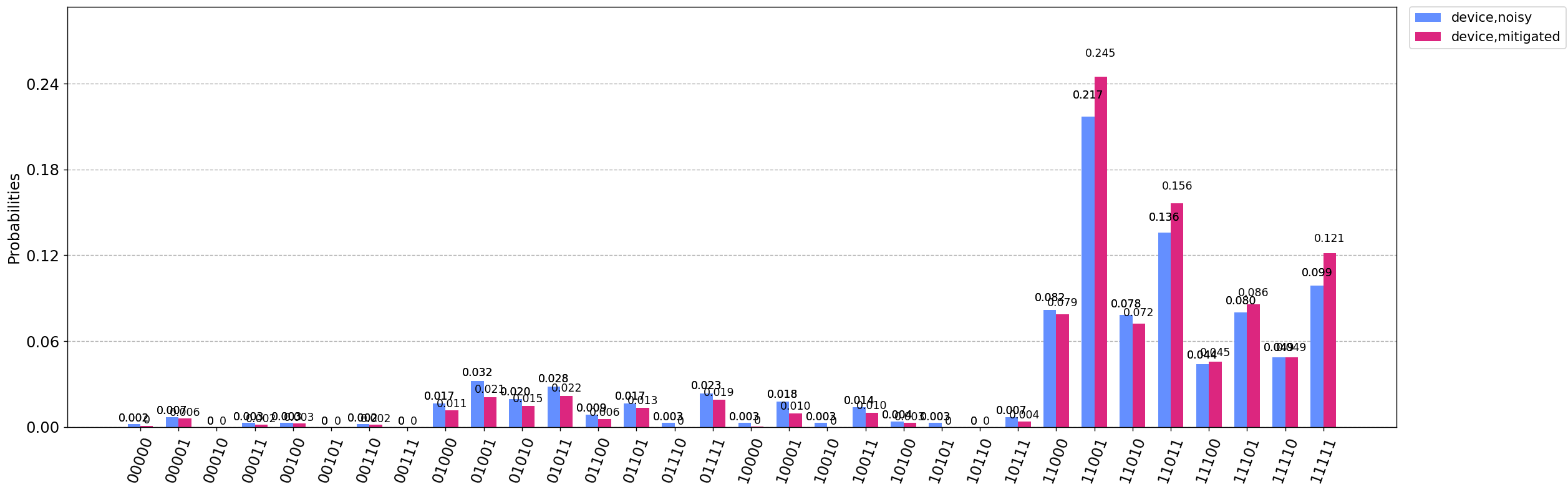}
    \caption{Results for implementation of an eavesdropper's attack on BB84 protocol using 5 qubits on IBM QX (ibmq\_manila) with error mitigation. The result of the Fig.\ref{SPAttackhist} is also plotted simultaneously here. The mitigated result is more accurate.}
    \label{SPAttackhistMit}
\end{figure}
The result of simulation for each attacked qubit is shown in Table.\ref{qkdtable6}. We compare those result with the result for the normal case in the Table.\ref{qkdtable5}. For q[0], the probability of getting the initial bit 1 decreases from 96.1\% to 69.6\% due to the attack. q[4] is not a part of key bit. Attack on q[4] doesn't affect our protocol. For q[3], though the eavesdropper decode in the same basis as Alice uses to encode, the probability of getting the initial value 1 decreases from 97.0\% to 88.6\%. It is probably due to gate operational error in the real quantum device as more gates operate in the circuit with attack scenario. 
\begin{table}[ht]
    \centering
    \begin{tabular}{|c|c|c|c|}
    \hline
    \hline
    Attacked qubit Index & q[0] & q[3] & q[4] \\[0.5ex]
    \hline
    Alice's qubit & 1 & 1 & 0  \\  values & & &  \\[0.5ex]
    \hline
    P(0) & 30.4\% & 11.4\% & 16.4\% \\[0.5ex]
    \hline
    P(1) & 69.6\% & 88.6\% & 83.6\% \\[0.5ex]
    \hline
    \end{tabular}
    \caption{These are the experimental results based on the BB84 protocol using four bases in an eavesdropping scenario. Probability of getting 0 \& 1 i.e. P(0) \& P(1) are shown for the three attacked qubits q[0], q[3] \& q[4] are shown here. }
    \label{qkdtable6}
\end{table}

We now calculate the fidelity ( as in section \ref{SecIIAiii}) between Alice's prepared state and Bob's revealed state while eavesdropping is taking place and then compare that fidelity with the case when no eavesdropping is happening. Comparison are shown in Table.\ref{Fid_table_attack}. Here only q[0], q[3], q[4] are subjected to attack, so fidelity calculation is done for only these qubits. 

\begin{table}[ht]
    \centering
    \begin{tabular}{|c|c|c|}
    \hline
    \hline
    Qubit Index & Fidelity without & Fidelity with \\ & eavesdropping & eavesdropping \\ [0.5ex]
    \hline
     q[0] & 0.9903666531616115  & 0.8727970305632149 \\ [0.5ex]
    \hline
     q[3] & 0.9988745629501058  & 0.9890826260851061 \\ [0.5ex]
    \hline
    q[4] & 0.21158200569024022 & 0.23464157714059153 \\[0.5ex]
    \hline
    \end{tabular}
    \caption{These are the Fidelity of the qubit states calculated based on the experimental results from ibmq\_manila of the BB84 protocol using four bases in normal and an eavesdropping scenario.}
    \label{Fid_table_attack}
\end{table}
q[0] and q[3] are accepted key bit. As the eavesdropper tries to measure q[0] with different decoding gates, the quantum state gets disturbed. As a result, fidelity decreases (as shown in Table:\ref{Fid_table_attack}). But as the eavesdropper measures q[3] with the same decoding gates, the attack remains invisible to Bob. q[4] is not a part of the accepted key bit so fidelity remains low.  

\subsubsection{Noise attack}
\label{SecIIBii}

\begin{figure}[H]
    \centering
    \includegraphics[scale=0.3]{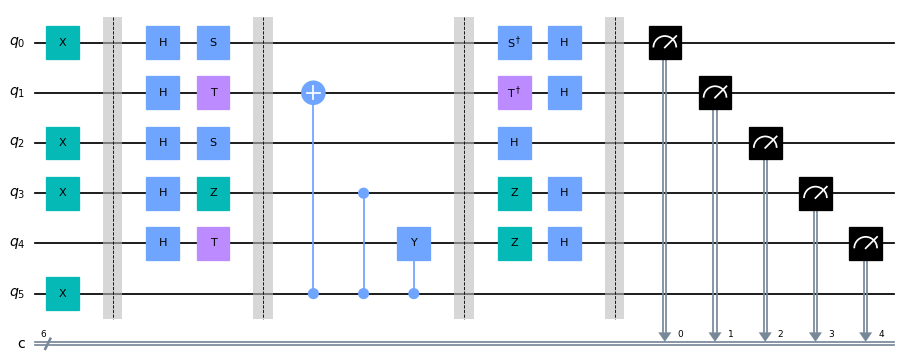}
    \caption{Noise attacks are implemented.q[1],q[3],q[4] are subjected to  CX, CZ \& CY noise attack respectively as shown.The control qubit q[5] is set high to show the effect.}
    \label{SPNoiAttack}
\end{figure}

\begin{figure}[H]
    \centering
    \includegraphics[scale=0.2]{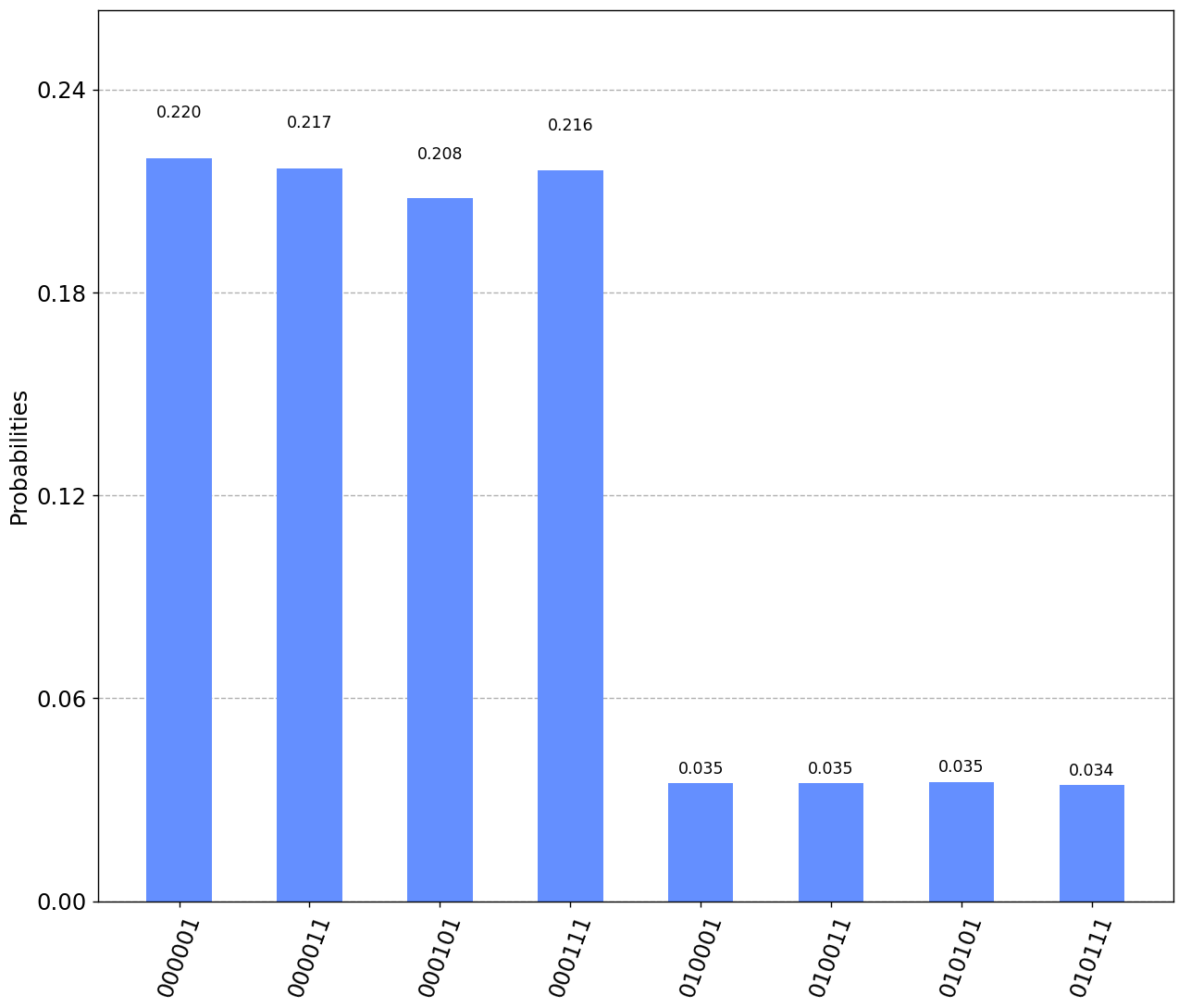}
    \caption{Results for implementation of noise attack on BB84 protocol using 6 qubits on IBM QX (ibmq\_qasm\_simulator). Once Bob reveals his measurements to Alice, they would realize that some attack has happened as the outputs do not tally with what Alice had encoded. The probability outcomes for all three attacked qubits are shown in Table.\ref{qkdtable6}.}
    \label{SPNoiAttackhist}
\end{figure}

We now can think of some noise occurrence in main information while data transmitting through a quantum channel. We can implement this type of noise with the help of some 2-qubit entangled gates like CX, CZ, and CY gates. If the control bit is 0, then these gates will not show any effect. So, we have shown the result taking the control bit as 1. The circuit is shown in Fig.\ref{SPNoiAttack} whereas the results of the implementation are in Fig.\ref{SPNoiAttackhist}. Table.\ref{qkdtable7} shows the probability of occurrence of each qubit for the noise attack. Evidently, probability of occurrence changes for those particular qubits due to attack. 

\begin{table}[ht]
    \centering
    \begin{tabular}{|c|c|c|c|}
    \hline
    \hline
    Qubit Index & q[1] & q[3] & q[4] \\[0.5ex]
    \hline
    Attack on Alice's & 0 & 1 & 0  \\ qubits (values) & & &  \\[0.5ex]
    \hline
    Type of Attack & CX & CZ & CY \\[0.5ex]
    \hline
    P(0) & 53.3\% & 97.7\% & 87.2\% \\[0.5ex]
    \hline
    P(1) & 46.7\% & 2.3\% & 12.8\% \\[0.5ex]
    \hline
    \end{tabular}
    \caption{These are the experimental results based on the BB84 protocol using four bases in a noise attack scenario. Parentheses include the probability of that particular bit.}
    \label{qkdtable7}
\end{table}
Also, we have calculated the fidelity of each qubit, subjected to noise attack, to show the change of qubit state due to the attack. Fidelity of q[1], q[3] and q[4] has been calculated (like in the section.\ref{SecIIAiii}) and tabulated in Table.\ref{Fid_table_nois}.
\begin{table}[ht]
    \centering
    \begin{tabular}{|c|c|c|}
    \hline
    \hline
    Qubit Index & Fidelity without & Fidelity with \\(Attack-type) & eavesdropping & eavesdropping \\ [0.5ex]
    \hline
     q[1] (CX) & 0.9998826058435594 & 0.5771227257614313 \\ [0.5ex]
    \hline
     q[3] (CZ)& 0.9988745629501058  &  0.07043737053093656\\ [0.5ex]
    \hline
    q[4] (CY)& 0.21158200569024022 & 0.8664884620665283 \\[0.5ex]
    \hline
    \end{tabular}
    \caption{These are the Fidelity of the qubit states calculated based on the experimental results from ibmq\_manila of the BB84 protocol using four bases in normal and an noise attack scenario.}
    \label{Fid_table_nois}
\end{table}

\begin{figure}[H]
    \centering
    \includegraphics[scale=0.35]{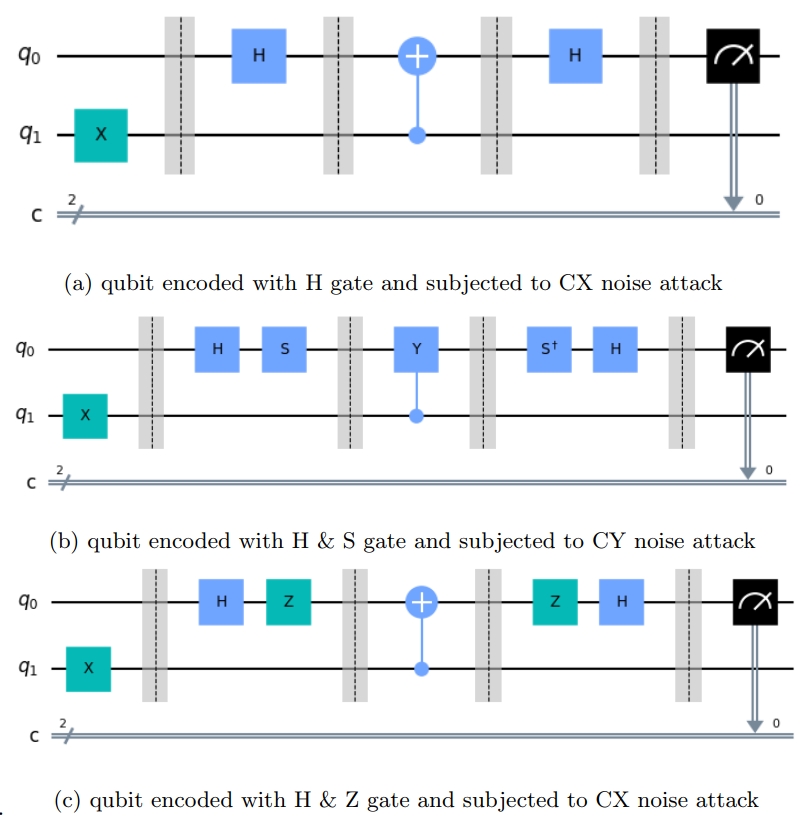}
    \caption{Circuits for the three fault-tolerant cases encoded in a certain basis and subjected to certain noise attacks. The initial qubit value is taken as 0 in each case.}
    \label{Exp_noise}
\end{figure}

As evident from Table.\ref{Fid_table_nois}, accepted key bits (here q[1], q[3]) are affected by the noise attack. The fidelity becomes very low for these qubits. So, in practical implementation, Bob will not get the expected result as the quantum state is changing considerably due to these noise attacks. Though there are few exceptions where certain noise attack doesn't affect the quantum state encoded in certain basis. Exceptions are shown in Fig.\ref{Exp_noise} and fidelity results are shown in Table.\ref{NoiseExp_tab}. As an explanation, for case (b), applying S gate after H gate leaves the qubit state through the z-axis of blotch sphere\cite{bloch} with a phase of $\pi/2$ to the $\Ket{1}$ state. Applying Y gate rotates the state at an angle of $\pi$ which leaves the qubit through the z-axis in the bloch sphere as before. Similarly, in the case of (a) \& (b), applying X gate doesn't change the qubit state through the z-axis in the bloch sphere. 
\begin{table}[ht]
    \centering
    \begin{tabular}{|c|c|c|}
    \hline
    \hline
    Qubit & Fidelity without & Fidelity with \\ & noise attack & noise attack \\ [0.5ex]
    \hline
     q[0] in (a) & 0.9903666531616115  & 0.9944090977516634 \\ [0.5ex]
    \hline
     q[0] in (b)& 0.9988745629501058  & 0.9937805250263796 \\ [0.5ex]
    \hline
    q[0] in (c) & 0.9975738546501054 & 0.9976180022158906 \\[0.5ex]
    \hline
    \end{tabular}
    \caption{These are the Fidelity of the qubit states of the circuits of Fig:
    \ref{Exp_noise} calculated based on the experimental results from ibmq\_manila. As evident, quantum state doesn't get changed much as fidelity remains very close to 1.}
    \label{NoiseExp_tab}
\end{table}

\section{SARG04 PROTOCOL}
\label{SecIII}
SARG04 protocol was built intending to create a more robust version of the BB84 protocol. The same four states of BB84 were used in the case of SARG04 but with a different information encoding process. In BB84 protocol, when laser pulse with lower-level attenuation is used to encode bit in a photon. If any pulse contains more than one photon, the eavesdropper can transmit a single photon to the receiver after splitting off the extra photon. This is a photon number splitting attack \cite{2000PhRvL..85.1330B}. SARG04 was defined by Scarani et al. in 2004 in Physical Review Letters as more efficient against PNS attack but it is equivalent to BB84 when viewed at the level of quantum processing \cite{2004PhRvL..92e7901S}.
The basic working of SARG04 protocol is described below:
\begin{itemize}
    \item \raggedright Alice intends to communicate with Bob. She starts with two strings of bit, x \& y. Each string is n bits long.
    \item Now she creates a state by encoding the two-bit string such that $y_i$(the ith bit in y string) decides on which basis $x_i$ is encoded. Value of $y_i$ 1 leads $x_i$ to be encoded in the hadamard basis. Otherwise, $x_i$ is encoded in the computational basis. 
    \item Alice sends her quantum state through a quantum channel. Bob receives the state. Eavesdropper, Eve also can have access to the state. But bit string y is only known to Alice, so, neither Bob nor Eve can distinguish the state of the qubit. 
    \item Bob creates a random bit-string $y'$ which is also n bits long. He decodes and measures the qubits such that $y'_i$ decides on which basis $x_i$ will be measured. For each qubit, Alice chooses one computational basis state and one hadamard basis state. To get the secret bits, Bob has to distinguish between the two states. For each qubit sent, Bob can check if the result is consistent with either possible state. If the measurement is consistent, then he cannot decide the exact encoding basis. But if the measurement is inconsistent, Bob can then deduce the exact state as well as the secret bit.\label{Process}
 
    \item From the remaining deduced bits, Alice shares a small randomly chosen part publicly with Bob and checks whether they agree to a certain number. If the check pass, they proceed to continue otherwise they drop that channel and start over. 
\end{itemize}
In our experimental realization, the scenario is much more simpler. As we typically run our circuit many times (8192 runs) and produce a histogram, we don't need that process\ref{Process} to decide the exact state. We can decide the exact state from the probability values of $\ket{0} \& \ket{1}$ in the produced histogram.
In simpler BB84 protocol, Alice announces the chosen basis for each qubit.  The advantage of SARG04 is that Alice doesn't announce the basis of the bits. So, to determine the state, Eve needs to store more copies of the qubit. We have experimentally shown the scenario in this section with the help of IBM QX.
\subsection{IMPLEMENTATION OF SARG04 PROTOCOLS ON IBM QX}
\label{SecIIIA}
We have simulated a 9-qubit SARG04 circuit using a ibmq\_qasm\_simulator. The first 3 qubits are the information bits, the next 3 qubits are Alice's reference bits and the last 3 qubits are Bob's reference bits. 
\begin{figure}[H]
    \centering
    \includegraphics[scale=0.35]{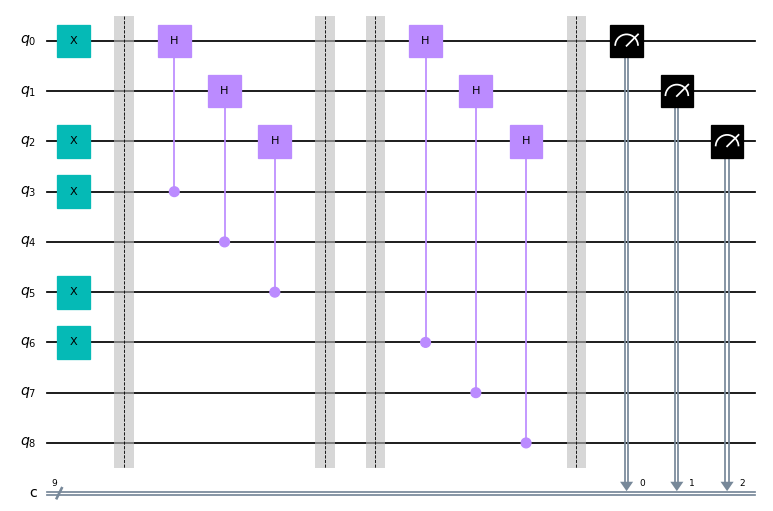}
    \caption{Implementation of SARG04 protocol with 3 information bits. The left part of the first barrier portraits the initial bit-strings. The part between first \& second barrier and the part between second \& third barrier portraits Alice's encoding and Bob's decoding respectively. The distance between the second and third barrier can be considered as the quantum channel. The right part of fourth barrier contains Bob's measurements.}
    \label{SARGcircuit}
\end{figure}
\begin{table}[ht]
    \centering
    \begin{tabular}{|c|c|c|c|}
    \hline
    \hline
    qubit index & q[0] & q[1] & q[2] \\[0.5ex]
    \hline
    Alice's reference bit & 1 & 0 & 1 \\[0.5ex]
    \hline
    Bob's reference bit & 1 & 0 & 0 \\[0.5ex]
    \hline
    Result & A & A & D \\[0.5ex]
    \hline
    \end{tabular}
    \caption{These are the Alice and Bob's reference bits for different information qubits. If reference bit is same for a qubit then the qubit is accepted for creating key bit, otherwise discarded.}
    \label{qkdtable8}
\end{table}
\begin{table}[ht]
    \centering
    \begin{tabular}{|c|c|}
    \hline\hline
    Qubit Index & Expected\\ & Result  \\ [0.5ex]
    \hline
    q[0] & 1 (100\%)\\[0.5ex]
    \hline
    q[1] & 0 (100\%) \\[0.5ex]
    \hline
    q[2] & 0 (50\%) \& 1 (50\%) \\[0.5ex]
    \hline
    \end{tabular}
    \caption{These are the expected results based on the SARG04 protocol. Parentheses include the probability of that particular bit.}
    \label{qkdtable9}
\end{table}
\subsubsection{Simulation using Local Simulator}
\label{SecIIIAi}
\begin{figure}[H]
    \centering
    \includegraphics[scale=0.13]{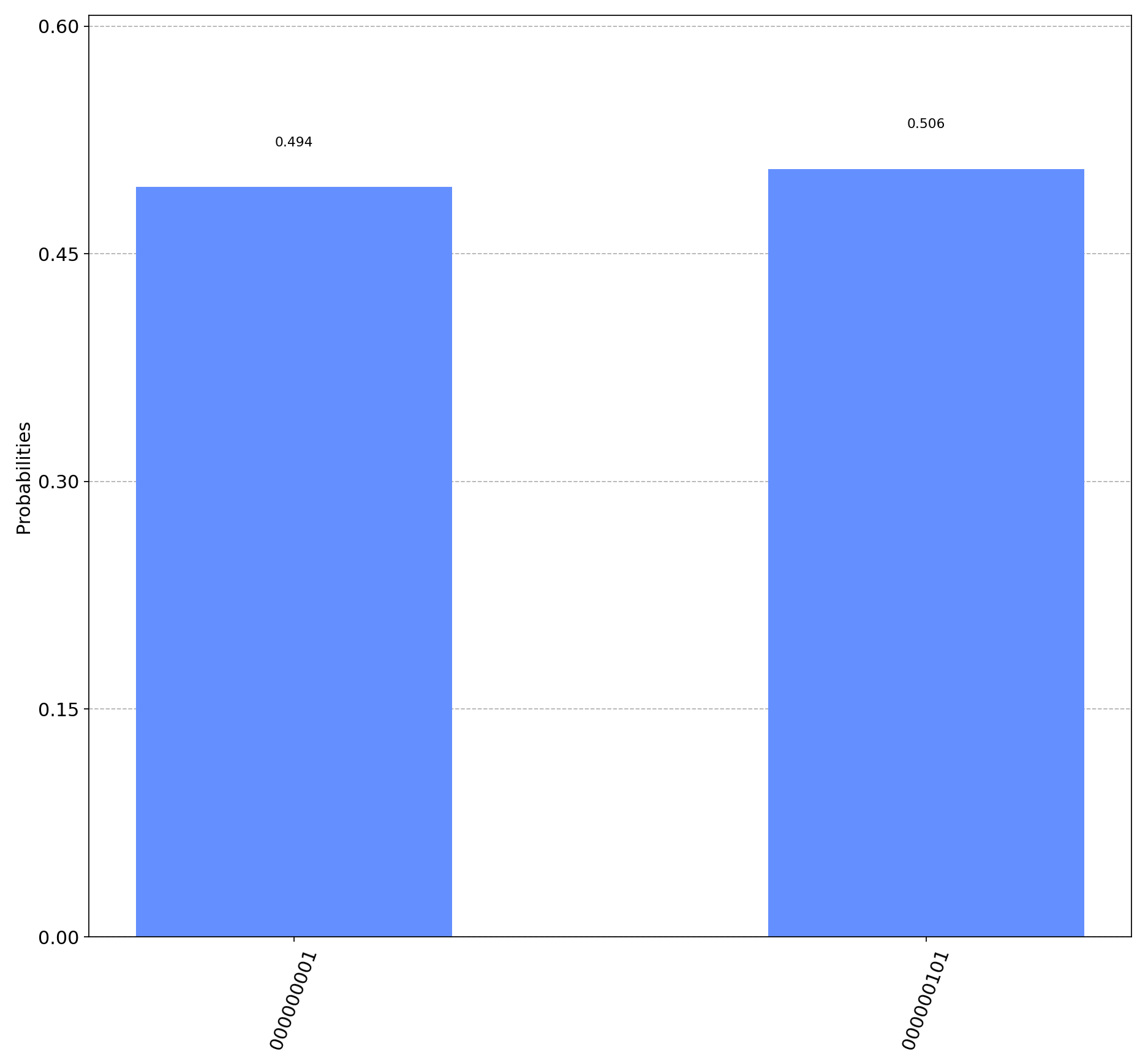}
    \caption{Result of implementation of SARG04 protocol acquired through IBM QX (ibmq\_qasm\_simulator) using 8192 runs. The three bits from the extreme right are information bits that were measured. In all of the outputs, the value of q[0] is 1 \& q[1] is 0 which implies that the probability of acquiring 1 \& 0 is respectively 100\% which is what we expected. Similarly, the probability of measuring q[2] and acquiring 0 is 49.4\% and that of 1 is 50.6\%. The results are calculated like this and tabulated in Table.\ref{qkdtable10}.}
    \label{SARGhist}
\end{figure}

Now firstly we simulate the implemented circuit in Fig.\ref{SARGcircuit} through a local simulator. As the simulator is not a real device, the histogram obtained , shown in Fig.\ref{SARGhist}, actually indicates the theoretically calculated result as in Table.\label{qkdtable9}. 
\subsubsection{Simulation using Real Quantum device(ibmq\_manila)}
\label{SecIIIAii}
We couldn't simulate the whole 9-qubit circuit of the Fig.\ref{SARGcircuit} due to our unavailability of real quantum processor of such qubit-volume. However, we simulated every qubit separately through ibmq\_manila and shown the result in Table.\label{qkdtable10}. The effect of noise in a real quantum device is evident from the comparison of experimental results in Table.\label{qkdtable10} and theoretically calculated result in Table.\label{qkdtable9}.
\begin{table}[ht]
    \centering
    \begin{tabular}{|c|c|c|}
    \hline\hline
    Qubit Index & Experimental\\ & Result  \\ [0.5ex]
    \hline
    q[0] & 0 (3.1\%) \& 1 (96.9\%)\\[0.5ex]
    \hline
    q[1] & 0 (99.1\%) \& 1 (0.9\%)\\[0.5ex]
    \hline
    q[2] & 0 (53.6\%) \& 1 (46.4\%) \\[0.5ex]
    \hline
    \end{tabular}
    \caption{These are the experimental results obtained by simulating each qubit of SARG04 protocol in IBM QX (ibmq\_manila) using 8192 runs. Parentheses include the probability of that particular bit.}
    \label{qkdtable10}
\end{table}
\subsubsection{Fidelity Calculation}
\label{SecIIIAiii}

We now calculate the fidelity(as in section.\ref{SecIIAiii}) between Alice's prepared state and Bob's revealed state in the SARG04 circuit. The required circuit simulations are done through IBM QX (ibmq\_manila) and the results are tabulated below.

\begin{table}[ht]
    \centering
    \begin{tabular}{|c|c|}
    \hline
    \hline
    Qubit Index & Fidelity \\[0.5ex]
    \hline
    q[0] & 0.9973847010664143  \\[0.5ex]
    \hline
    q[1] & 0.9982795399056618 \\[0.5ex]
    \hline
    q[2] & 0.47984599999999983 \\[0.5ex]
    \hline
    \end{tabular}
    \caption{Fidelity values for individual qubits of SARG04 protocol. As expected from Table.\ref{qkdtable7}, q[0] and q[1] have fidelity close to 1 as encoding and decoding reference bit is same for these. q[2] has low fidelity so discarded.}
    \label{fid_table_SARG}
\end{table}

\subsection{DIFFERENT ATTACKS ON THE QUANTUM CIRCUIT}
\label{SecIIIB}

\subsubsection{Third party interpretation}
\label{SecIIIBi}
\begin{figure}[H]
    \centering
    \includegraphics[scale=0.45]{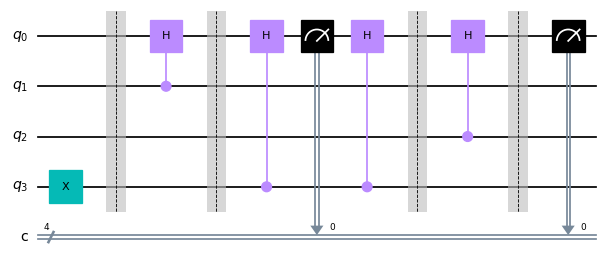}
    \caption{Eavesdropper tries to measure the state in hadamard basis as her reference bit is 1. Whereas reference bit of Alice and Bob are both 0.  }
    \label{SARGattack}
\end{figure}

We now implement an eavesdropper attack on the SARG04 circuit. For that, we consider only q[1] from the circuit of Fig.\ref{SARGcircuit}. Eavesdropper decodes the state in a basis depending on the reference bit, randomly chosen by her. So, the eavesdropper has a probability of 0.5 of choosing the right basis for measurement but she is never sure as Alice never shares her encoding basis. In our case, we have shown an eavesdropper attack with wrong decoding basis Fig.\ref{SARGattack} to show the change in the quantum state. It is obvious from Table.\ref{Fid_SARG_attack}, as fidelity decreases considerably in the case of eavesdropping. 

\begin{table}[ht]
    \centering
    \begin{tabular}{|c|c|c|}
    \hline
    \hline
    Qubit Index & Fidelity without & Fidelity with \\ & eavesdropping & eavesdropping \\ [0.5ex]
    \hline
     q[1] & 0.9982795399056618  & 0.5998332266675055 \\ [0.5ex]
    \hline
    \end{tabular}
    \caption{This is the comparison of fidelity of the qubit state in normal and an eavesdropping case, calculated based on the experimental results from ibmq\_manila of q[1] of the SARG04 protocol.}
    \label{Fid_SARG_attack}
\end{table}
\subsubsection{Noise attack}
\label{SecIIIBii}
Now we can show the effect of noise in the quantum channel with the help of some 2-qubit entangled gates like CX, CZ, and CY gates as done before for the BB84 protocol in this paper (section.\ref{SecIIBii}). We have already seen the effect of these noise attacks when a qubit is encoded and decoded in hadamard basis. The quantum state will change considerably for CY \& CZ attack but there will be no effect in the quantum state in case of CX attack(Fig.\ref{Exp_noise}, Table.\ref{NoiseExp_tab}). For the case of computational basis, CX \& CY attack will change the state very significantly as X \& Y gates are responsible for bit-flip. CZ attack will not have any effect on computational basis. 
\section{CONCLUSION}
\label{SecIV}

Here, implementation of BB84 protocol (with four bases) and SARG04 protocol along with eavesdropping case on IBM Quantum Experience's platform (ibmq\_manila and ibmq\_qasm\_simulator) was carried out. We then compared the expected and acquired experimental values. The values were compared and the overlap was found acceptable. The implementations help us visualize the theoretical concept by employing gate operations, thus providing an insight into the fundamentals of quantum computing.

\section{ACKNOWLEDGEMENTS}
\label{SecV}
S.G. and H.M would like to thank IISER Kolkata for providing hospitality during the course of this project. B.K.B. acknowledges the support of Institute Fellowship provided by IISER Kolkata. The authors acknowledge the support of IBM Quantum Experience for producing experimental results. The views expressed are those of the authors and do not reflect the official policy of IBM or IBM Quantum Experience team.


\begin{thebibliography}{10}
\label{SecRef}
\expandafter\ifx\csname url\endcsname\relax
\def\url#1{\texttt{#1}}\fi
\expandafter\ifx\csname urlprefix\endcsname\relax\def\urlprefix{URL }\fi
\providecommand{\bibinfo}[2]{#2}
\providecommand{\eprint}[2][]{\url{#2}}

\bibitem{RQC2011Science2011} R. Hughes and J. Nordholt, Refining Quantum Cryptography, Science \textbf{333}, 1584 (2011).

\bibitem{qkd_PathakCRC2018} A. Pathak, Elements of Quantum Computation and Quantum Communication, CRC Press (2018).

\bibitem{BBQCIEEE1984} C. H. Bennett and G. Brassard, Quantum cryptography: Public key distribution and coin tossing, IEEE International Conference on Computers, Systems and Signal Processing, \textbf{175}, 8 (1984).

\bibitem{BCBTCS2014} C.H. Bennett, H. Charles, G. Brassard, Quantum cryptography: Public key distribution and coin tossing, Theoretical Computer Science, \textbf{560}, Part 1: 7–11, (2014).

\bibitem{QKDL12} D. Bouwmeester, A. Ekert, A. Zeilinger, Lecture 12: Quantum key distribution, The Physics of Quantum Information.

\bibitem{NCCUP2000} M. Nielson, I. Chuang, Quantum Computing and Quantum Information, Cambridge University Press, (2000).

\bibitem{qkd_WoottersNature1982} W.~K. Wootters and W.~H. Zurek, A Single Quantum Cannot be Cloned, Nature \textbf{299}, 802 (1982).

\bibitem{MKAAANICCAIS2019} D. AL-Mubayedh, M. AL-Khalis, G. AL-Azman, M. AL-Abdali, M. AlFosail, and N. Nagy, Quantum Cryptography on IBM QX, International Conference on Computer Applications \& Information Security, (2019). 

\bibitem{HFPR2014} H. Fan, Y. Wang, L. Jing, J. Yue, H. Shi, Y. Zhang, L. Mu, Quantum cloning machines and the applications, Phys. Rep. \textbf{544}, 241-322 (2014)

\bibitem{BFO2017} B. Fröhlich, M. Lucamarini, J. Dynes, L. Comandar, W. Tam, A. Plews, A. Sharpe, Z. Yuan, and A. Shields, Long-distance quantum key distribution secure against coherent attacks, Optica, \textbf{4} (1), 163-167 (2017)

\bibitem{AFPRA2012} A. Ferenczi, V. Narasimhachar, and N. Lütkenhaus, Security proof of the unbalanced phase-encoded Bennett-Brassard 1984 protocol, Phys. Rev. A, \textbf{86} (2012)

\bibitem{CHBJoC1992} C. H Bennett \textit{et al.}, Experimental Quantum Cryptography, Journal of Cryptology, \textbf{5}, 3–28 (1992)

\bibitem{MMNP2019} M. Minder \textit{et al.}, Experimental quantum key distribution beyond the repeaterless secret key capacity, Nature Photonics, \textbf{13}, 334–338 (2019)

\bibitem{DAMIEEE2013} Dhoha AL-Mubayedh \textit{et al.}, Quantum Cryptography on IBM QX, IEEE Xplore, 18851058 (2013)

\bibitem{2004PhRvL..92e7901S} Valerio Scarani, Antonio Acín, Grégoire Ribordy, and Nicolas Gisin, Quantum Cryptography Protocols Robust against Photon Number Splitting Attacks for Weak Laser Pulse Implementations, Physical Review Letters,14995344 (2004)

\bibitem{Jozsa1994} R. Jozsa, Fidelity for Mixed Quantum States, Journal of Modern Optics, 23152323 (1994)

\bibitem{BKbehera}
Behera B.K., Banerjee A., Panigrahi P.K., Experimental realization of quantum cheque using a five-qubit quantum computer.
Quantum Inf Process 16, 312 (2017). https://doi.org/10.1007/s11128-017-1762-0

\bibitem{ERQKD2020} A. Warke, B.K.Behera and P.K. Panigrahi, Experimental realization of three quantum key distribution protocols, Quantum Inf Process 19, 407 (2020). https://doi.org/10.1007/s11128-020-02914-z

\bibitem{bloch} 
Single Qubit Gates.
https://qiskit.org/textbook/ch-states/representing-qubit-states.html

\bibitem{2000PhRvL..85.1330B} Brassard, Gilles, Lütkenhaus, Norbert,  Mor, Tal, Sanders, Barry C., Limitations on Practical Quantum Cryptography, Physical Review Letters,(2000)

\bibitem{GEM2020} Manpreet Singh Jattana, Fengping Jin, Hans De Raedt \& Kristel Michielsen, General error mitigation for quantum circuits. Quantum Inf Process 19, 414 (2020). https://doi.org/10.1007/s11128-020-02913-0


\end{thebibliography}
\end{document}

\title{Supplementary Information: Demonstration of a general fault-tolerant quantum error detection code for $(2n+1)$-qubit entangled state on IBM 16-qubit quantum computer}

\author{Ranveer Kumar Singh}
\email{ranveersfl@gmail.com}
\affiliation{Department of Mathematics, \\Indian Institute of Science Education and Research Bhopal, Bhauri 462066, Madhya Pradesh, India}
\author{Bishvanwesha Panda}
\email{bishvanweshapanda@gmail.com}
\affiliation{Indian Institute of Science Education and Research Kolkata,\\ Mohanpur 741246, West Bengal, India}

\author{Bikash K. Behera}
\email{bkb18rs025@iiserkol.ac.in}
\author{Prasanta K. Panigrahi}
\email{pprasanta@iiserkol.ac.in}
\affiliation{Department of Physical Sciences,\\ Indian Institute of Science Education and Research Kolkata, Mohanpur 741246, West Bengal, India}

\maketitle

\section{Simulation of error detection protocol}
For simulating the error detection protocol, we used QISKit to take both simulation results. The QASM code for the same is as follows: 

\lstinputlisting[language=Python]

\subsection{Measurement data}
We performed all the simulations on QISKit and recorded the countings of each of the measurement result over the two ancillary error syndrome qubit in 8192 shots. From the countings, the probability of each error \textit{i.e.} bit-flip error, phase-flip error and arbitrary phase-change error was extracted. The data is shown in the table \ref{qed_table1} below.
\begin{table}[h!]
\begin{center}
 \begin{tabular}{c c c c c} 
 \hline
 \hline
 Error & $\{0 , +\}$ & $\{1 , +\}$&$\{0 , -\}$&$\{1 , -\}$ \\ [0.5ex] 
 \hline
 \hline
 $Y_{\pi/3}$ &0.747 &0 &0&0.253   \\ 
 \hline
 $X_{\pi/3}$ &0.75 &0.25&0&0  \\
 \hline
 $X_{\pi/3}Y_{\pi/3}$ & 0.56&0.185&0.066&0.188\\
 \hline
 $X_{\pi/3}Y_{2\pi/3}$ & 0.18&0.063&0.184&0.574   \\  
\hline
 $X_{2\pi/3}Y_{\pi/3}$ & 0.19&0.55&0.195&0.063   \\ 
 \hline
 $X_{2\pi/3}Y_{2\pi/3}$ & 0.06&0.19&0.056&0.185 \\
 \hline
  $R=X_{\pi/2}Y_{\pi/2}$&0.25&0.252&0.252&0.245\\
  \hline
  $H$&0&0.503&0.497&0\\[1ex] 
 \hline
 \hline
\end{tabular}
\caption{\textbf{Probability of each type of error.} Here $\{0,+\},\{1,+\},\{0,-\}$ and $\{1,-\}$ represent the two qubit states $\Ket{00},\Ket{10},\Ket{01}$ and $\Ket{11}$ respectively. $+$ is the shorthand for $\Ket{+}=\frac{1}{\sqrt{2}}\big(\Ket{0}+\Ket{1}\big)$ and $-$ is the shorthand for $\Ket{-}=\frac{1}{\sqrt{2}}\big(\Ket{0}-\Ket{1}\big)$. $\Ket{+},\Ket{-}$ are the states of the second ancillary syndrome qubit before the Hadamard operation in the circuit of Fig. \ref{qed_fig2} in the bit-flip and phase-flip cases respectively.}
\label{qed_table1}
\end{center}
\end{table}
\begin{table}[h!]
\begin{center}
 \begin{tabular}{c c c c c} 
 \hline
 \hline
 $\theta$ & $\{0 , +\}$ & $\{1 , +\}$&$\{0 , -\}$&$\{1 , -\}$ \\ [0.5ex] 
 \hline
 \hline
 $-\pi$ &0 &1 &0&0   \\ 
 \hline
 $-14\pi/15$ &0.012 &0.988&0&0  \\
 \hline
 $-13\pi/15$ & 0.045&0.955&0&0\\
 \hline
 $-12\pi/15$ & 0.092&0.908&0&0   \\  
\hline
 $-11\pi/15$ & 0.1644&0.8356&0&0.   \\ 
 \hline
 $-10\pi/15$ & 0.25&0.75&0&0 \\
 \hline
  $-9\pi/15$&0.35&0.65&0&0\\
  \hline
  $-8\pi/15$&0.45&0.55&0&0\\
  \hline
  $-7\pi/15$&0.55&0.45&0&0\\
  \hline
   $-6\pi/15$&0.65&0.35&0&0\\
  \hline
  $-5\pi/15$&0.752&0.248&0&0\\
  \hline
 $-4\pi/15$&0.843&0.157&0&0\\
  \hline
   $-3\pi/15$&0.905&0.095&0&0\\
  \hline
  $-2\pi/15$&0.952&0.048&0&0\\
  \hline
  $-\pi/15$&0.987&0.013&0&0\\
  \hline
  $0$&1&0&0&0\\
  \hline
  $\pi/15$&0.99&0.0091&0&0\\
  \hline
 $2\pi/15$&0.957&0.043&0&0\\
  [1ex] 
 \hline
 \hline
\end{tabular}
\caption{\textbf{Probability of each type of error for applied error $\varepsilon=X_{\theta}$ with varying $\theta$.}}
\label{qed_table2}
\end{center}
\end{table}
\begin{table}[h!]
\begin{center}
 \begin{tabular}{c c c c c} 
 \hline
 \hline
 $\theta$ & $\{0 , +\}$ & $\{1 , +\}$&$\{0 , -\}$&$\{1 , -\}$ \\ [0.5ex] 
 \hline
 \hline
 $3\pi/15$ &0.906 &0.094 &0&0   \\ 
 \hline
 $4\pi/15$ &0.831 &0.17&0&0  \\
 \hline
 $5\pi/15$ & 0.75&0.25&0&0\\
 \hline
 $6\pi/15$ & 0.658&0.342&0&0   \\  
\hline
 $7\pi/15$ & 0.554&0.446&0&0.   \\ 
 \hline
 $8\pi/15$ & 0.436&0.563&0&0 \\
 \hline
  $9\pi/15$&0.34&0.66&0&0\\
  \hline
  $10\pi/15$&0.25&0.75&0&0\\
  \hline
  $11\pi/15$&0.164&0.836&0&0\\
  \hline
   $12\pi/15$&0.094&0.91&0&0\\
  \hline
  $13\pi/15$&0.044&0.956&0&0\\
  \hline
  $14\pi/15$&0.012&0.988&0&0\\
  \hline
  $\pi$&0&1&0&0\\[1ex]
  \hline
 \hline
\end{tabular}
\caption{\textbf{Probability of each type of error for applied error $\varepsilon=X_{\theta}$ with varying $\theta$.} (continued\dots)}
\label{qed_table3}
\end{center}
\end{table}
\begin{table}[h!]
\begin{center}
 \begin{tabular}{c c c c c} 
 \hline
 \hline
 $\theta$ & $\{0 , +\}$ & $\{1 , +\}$&$\{0 , -\}$&$\{1 , -\}$ \\ [0.5ex] 
 \hline
 \hline
 $-\pi$ &0 &0 &0&1   \\ 
 \hline
 $-14\pi/15$ &0.011 &0&0&0.99  \\
 \hline
 $-13\pi/15$ & 0.044&0&0&0.956\\
 \hline
 $-12\pi/15$ & 0.098&0&0&0.902   \\  
\hline
 $-11\pi/15$ & 0.166&0&0&0.834   \\ 
 \hline
 $-10\pi/15$ & 0.251&0&0&0.75 \\
 \hline
  $-9\pi/15$&0.35&0&0&651\\
  \hline
  $-8\pi/15$&0.45&0&0&554\\
  \hline
  $-7\pi/15$&0.56&0&0&0.44\\
  \hline
  $-6\pi/15$&0.66&0&0&0.34\\
  \hline
  $-5\pi/15$&0.75&0&0&0.25\\
  \hline
 $-4\pi/15$&0.84&0&0&0.164\\
  \hline
  $-3\pi/15$&0.905&0&0&0.095\\
  \hline
  $-2\pi/15$&0.957&0&0&0.042\\
  \hline
  $-\pi/15$&0.988&0&0&0.012\\
  \hline
  $0$&1&0&0&0\\
  \hline
  $\pi/15$&0.989&0&0&0.011\\
  \hline
 $2\pi/15$&0.957&0&0&0.043\\
  [1ex] 
 \hline
 \hline
\end{tabular}
\caption{\textbf{Probability of each type of error for applied error $\varepsilon=Y_{\theta}$ with varying $\theta$.}}
\label{qed_table4}
\end{center}
\end{table}
\begin{table}[h!]
\begin{center}
 \begin{tabular}{c c c c c} 
 \hline
 \hline
 $\theta$ & $\{0 , +\}$ & $\{1 , +\}$&$\{0 , -\}$&$\{1 , -\}$ \\ [0.5ex] 
 \hline
 \hline
 $3\pi/15$ &0.905 &0 &0&0.095   \\ 
 \hline
 $4\pi/15$ &0.831 &0&0&0.17  \\
 \hline
 $5\pi/15$ & 0.751&0&0&0.25\\
 \hline
 $6\pi/15$ & 0.65&0&0&0.35   \\  
\hline
 $7\pi/15$ & 0.56&0&0&0.44   \\ 
 \hline
 $8\pi/15$ & 0.45&0&0&0.552 \\
 \hline
  $9\pi/15$&0.35&0&0&0.65\\
  \hline
  $10\pi/15$&0.25&0&0&0.75\\
  \hline
  $11\pi/15$&0.168&0&0&0.832\\
  \hline
  $12\pi/15$&0.092&0&0&0.908\\
  \hline
  $13\pi/15$&0.039&0&0&0.96\\
  \hline
 $14\pi/15$&0.012&0&0&0.99\\
  \hline
  $\pi$&0&0&0&1\\[1ex]
  \hline
 \hline
\end{tabular}
\caption{\textbf{Probability of each type of error for applied error $\varepsilon=Y_{\theta}$ with varying $\theta$.} (continued\dots) }
\label{qed_table5}
\end{center}
\end{table}
\begin{table}[h!]
\begin{center}
 \begin{tabular}{c c c c c} 
 \hline
 \hline
 $\theta$ & $\{0 , +\}$ & $\{1 , +\}$&$\{0 , -\}$&$\{1 , -\}$ \\ [0.5ex] 
 \hline
 \hline
 $-\pi$ &0 &0 &1&0   \\ 
 \hline
 $-14\pi/15$ &0.011 &0&0.988&0  \\
 \hline
 $-13\pi/15$ & 0.044&0&0.956&0\\
 \hline
 $-12\pi/15$ & 0.096&0&0.904&0   \\  
\hline
 $-11\pi/15$ & 0.163&0&0.837&0   \\ 
 \hline
 $-10\pi/15$ & 0.25&0&0.75&0 \\
 \hline
  $-9\pi/15$&0.35&0&0.65&0\\
  \hline
  $-8\pi/15$&0.45&0&0.55&0\\
  \hline
  $-7\pi/15$&0.55&0&0.45&0\\
  \hline
  $-6\pi/15$&0.65&0&0.35&0\\
  \hline
  $-5\pi/15$&0.75&0&0.25&0\\
  \hline
  $-4\pi/15$&0.83&0&0.17&0\\
  \hline
  $-3\pi/15$&0.91&0&0.09&0\\
  \hline
  $-2\pi/15$&0.96&0&0.04&0\\
  \hline
  $-\pi/15$&0.99&0&0.011&0\\
  \hline
  $0$&1&0&0&0\\
  \hline
  $\pi/15$&0.99&0&0.01&0\\
  \hline
 $2\pi/15$&0.96&0&0.043&0\\
  [1ex] 
 \hline
 \hline
\end{tabular}
\caption{\textbf{Probability of each type of error for applied error $\varepsilon=Z_{\theta}$ with varying $\theta$.}}
\label{qed_table6}
\end{center}
\end{table}
\begin{table}[h!]
\begin{center}
 \begin{tabular}{c c c c c} 
 \hline
 \hline
 $\theta$ & $\{0 , +\}$ & $\{1 , +\}$&$\{0 , -\}$&$\{1 , -\}$ \\ [0.5ex] 
 \hline
 \hline
 $3\pi/15$ &0.903 &0 &0.097&0   \\ 
 \hline
 $4\pi/15$ &0.831 &0&0.17&0  \\
 \hline
 $5\pi/15$ & 0.76&0&0.24&0\\
 \hline
 $6\pi/15$ & 0.65&0&0.34&0   \\  
\hline
 $7\pi/15$ & 0.55&0&0.45&0   \\ 
 \hline
 $8\pi/15$ & 0.44&0&0.56&0 \\
 \hline
  $9\pi/15$&0.35&0&0.65&0\\
  \hline
  $10\pi/15$&0.25&0&0.75&0\\
  \hline
  $11\pi/15$&0.17&0&0.83&0\\
  \hline
  $12\pi/15$&0.099&0&0.9&0\\
  \hline
  $13\pi/15$&0.04&0&0.96&0\\
  \hline
$14\pi/15$&0.011&0&0.989&0\\
  \hline
  $\pi$&0&0&1&0\\[1ex]
  \hline
 \hline
\end{tabular}
\caption{\textbf{Probability of each type of error for applied error $\varepsilon=Z_{\theta}$ with varying $\theta$.} (continued\dots)}
\label{qed_table7}
\end{center}
\end{table}